\documentclass[a4paper,12pt]{article}
\pdfoutput=1
\usepackage[letterpaper]{geometry}
\usepackage{amsmath , amssymb, ascmac}
\usepackage{bm}
\usepackage{cancel}
\usepackage{comment}
\usepackage{mathtools}
\usepackage{float}
\usepackage{mathrsfs}
\usepackage{ytableau}
\usepackage[setpagesize=false]{hyperref}
\numberwithin{equation}{section}

\newcommand{\diff}{\mathrm{d}}

\newcommand{\del}[2]{\frac{\partial #1}{\partial #2}}

\newcommand{\newA}{\\ &\qquad \qquad}

\hypersetup{
 setpagesize=false,
 bookmarks=true,
 bookmarksdepth=tocdepth,
 bookmarksnumbered=true,
 colorlinks=false,
 hidelinks,
 pdftitle={},
 pdfsubject={},
 pdfauthor={},
 pdfkeywords={}}

\makeatletter
\renewcommand*{\eqref}[1]{
  \hyperref[{#1}]{\textup{\tagform@{\ref*{#1}}}}%
}
\makeatother

\begin{document}

\begin{titlepage}

\title{
\vspace{-1.5cm}\begin{flushright}
	{\normalsize TIT/HEP-693 \\ \vspace{-0.3 cm}March 2023}
\end{flushright}\vspace{3 cm}
Character Expansion Methods for $\mathrm{USp}(2N)$, $\mathrm{SO}(n)$, and $\mathrm{O}(n)$ using the Characters of the Symmetric Group }
\author{Akihiro Sei\thanks{E-mail: a.sei@th.phys.titech.ac.jp}\\\textit{\small Department of Physics, Tokyo Institute of Technology, Tokyo 152-8551, Japan}}
\date{}

\maketitle

\begin{abstract}
In theories with supersymmetry,  we can calculate a special partition function, known as the superconformal index. 
In particular, for a gauge group of $\mathrm{U}(N)$ and particles belonging to the adjoint representation, there is a fast method known as the character expansion method, which uses the characters of the symmetric group. 
In this paper, we extend this method to theories of particles belonging to specific representations of the gauge groups: $\mathrm{USp}(2N)$, $\mathrm{SO}(n)$, and $\mathrm{O}(n)$. 
Furthermore, we propose a formula, which gives the large $N$ limit without using the characters.
\end{abstract}

\tableofcontents
\clearpage
\ytableausetup{boxsize=1em}
\ytableausetup{centertableaux}
\ytableausetup{aligntableaux=center}

\end{titlepage}
\section{Introduction}\label{SecIntro}

In theories with supersymmetry, we can define a special partition function, known as the superconformal index \cite{Kinney:2005ej}. This function was calculated for supersymmetric theories in different space-time dimensions.
Such calculation is possible because the index is independent of the coupling constants and infinite dimensional path integral reduces to finite dimensional matrix integral.

Let us consider a theory of particles belonging to a representation $R$ of a gauge group $G$.
The partition function is expressed as
\begin{align}
	I_R^G (x) = \int [\diff U]_G \; \exp\left[ \sum_{m = 1}^{\infty} \frac{1}{m} f(x^m) \chi_{R}^{G} (U^m) \right],\label{indexeq}
\end{align}
where $[\diff U]_G$ is the Haar measure of $G$, normalized by $\int [\diff U]_G = 1$, $\chi_{R}^{G} (U)$ is the character of the representation $R$ of $G$, $f(x)$ is the single-particle partition function, defined by the trace over the single-particle states.
$x$ denotes a collection of fugacities.
If $x=(t,y,v,w)$ then $f(x^m) = f(t^m,y^m,v^m,w^m)$.
$\int [\diff U]_G$ is operation to extract only physical states, which belong to the trivial representation.
$m$ is the number of particles in the states.

One way to calculate \eqref{indexeq} is as follows. Firstly, we diagonalize $U$, express the integrand as a function of the eigenvalues of $U$, Tayler expand the integrand, and integrate it over the eigenvalues. However, this simple method is time-consuming, and to obtain the result up to high order of expansion we need efficient methods.
  
Let us rewrite the integrand as
\begin{align}
	\exp\left[ \sum_{m = 1}^{\infty} \frac{1}{m} f(x^m) \chi_{R}^{G} (U^m) \right]
		&= \sum_{\lambda} \left[\frac{ f_{\lambda}(x) }{z_{\lambda}} \prod_{i=1}^{l(\lambda)}\chi_{R}^{G}(U^{\lambda_i}) \right],\label{sumpar}
\end{align}
where $\lambda$ is an integer partition $(\lambda_1,\lambda_2,\dots), \;(\lambda_1 \ge \lambda_2 \ge \cdots \ge \lambda_{l(\lambda)} > 0)$ and $l(\lambda)$ is the length of $\lambda$. $\sum_{\lambda}$ is the summation over all partitions. We defined notations:
\begin{align}
	f_{\lambda} &\equiv \prod_{i = 1}^{l(\lambda)} f(x^{\lambda_i}), \\
	z_{\lambda} &\equiv \prod_{m=1}^{\infty} k_m ! \;m^{k_m},
\end{align}
where $k_m$ is the number of $m$ in $\lambda$. \ytableausetup{boxsize=0.4em}For example,
\begin{align}
	\lambda = \{3,2,1,1\} (= \ydiagram[]{3,2,1,1}\, ) \quad\Longrightarrow\quad l(\lambda) = 4,\  k_1 = 2,\ k_2 = 1,\ k_3 = 1.
\end{align}

In the case of $G = \mathrm{U}(N)$ and $R$ being the adjoint representation, we can calculate easily the integral $I_\mathrm{adj}^{\mathrm{U}(N)}$ by using the method ``the character expansion'' \cite{Dolan:2007rq, Dutta:2007ws}. 
We first review the method \cite{Dolan:2007rq, Dutta:2007ws} for $G = \mathrm{U}(N)$ (and $R$ being the adjoint representation) theory in Section \ref{SecU}. Next, in Section \ref{SecUSp}, \ref{SecSO} and \ref{SecO}, we will extend the methods to $\mathrm{USp} (2N)$, $\mathrm{SO} (2N + 1)$ and $\mathrm{SO} (2N)$, and $\mathrm{O} (2N + 1)$ and $\mathrm{O} (2N)$, respectively. In addition, in Section \ref{SecLargeN}, we will conjecture a formula for large $N$. 
Finally, we give an example in Appendix \ref{SecExample}.

\subsection{$\mathrm{U} (N)$ case}\label{SecU}
Let us consider the case of $G=\mathrm{U}(N)$.

If $R$ is the fundamental representation,  $I_{\mathrm{fund}}^{\mathrm{U}(N)} = 1$ because all (single) particle states have the positive $\mathrm{U}(1)$ charge and no states are neutral except the vacuum state.

Let us consider the case that $R$ is the adjoint representation \cite{Dolan:2007rq, Dutta:2007ws}.
$\chi_{\mathrm{adj}}^{\mathrm{U}(N)} (U) = (\mathrm{tr} U)(\mathrm{tr} U^\dagger)$ holds and the integral \eqref{indexeq} becomes
\begin{align}
	I_{\mathrm{adj}}^{\mathrm{U}(N)} (x) &= \sum_{\lambda} \left[\frac{ f_{\lambda}(x) }{z_{\lambda}} \int [\diff U]_{\mathrm{U} (N)} \,t_{\lambda} (U) \left(t_{\lambda} (U)\right)^* \right],
\end{align}
where $t_{\lambda} (U) \equiv \prod_{i=1}^{l(\lambda)} \mathrm{tr} (U^{\lambda_i})$.
We can expand $t_{\lambda} (U)$ into characters $\chi_{\mu}^{\mathrm{U}(N)} (U)$ by using the Frobenius formula
\begin{align}
	t_{\lambda} (U) \equiv \prod_{i=1}^{l(\lambda)} \mathrm{tr} (U^{\lambda_i}) &= \sum_{\mu \vdash |\lambda|, l(\mu) \le N} \chi_{\mu}^S (\lambda) \chi_{\mu}^{\mathrm{U} (N)} (U), \label{forbeEq}
\end{align}
where $\chi_{\mu}^S (\lambda)$ is the character of the symmetric group $S_{|\lambda|}$.
Here, we use partitions $\mu$ and $\lambda$ to specify an irreducible representation and a conjugacy class of the symmetric group $S_{|\lambda|}$, respectively. More details on this can be found in \cite{sym}. 
By using the orthonormality of characters 
\begin{align}
	\int [\diff U]_{\mathrm{U} (N)} \; \chi_{\mu}^{\mathrm{U} (N)} (U) \left(\chi_{\nu}^{\mathrm{U} (N)} (U)\right)^* &= \delta_{\mu\nu},
\end{align}
we obtain
\begin{align}
	I_{\mathrm{adj}}^{\mathrm{U}(N)} (x) &= \sum_{\lambda} \left[\frac{ f_{\lambda}(x) }{z_{\lambda}} \sum_{ \mu \vdash |\lambda|: l(\mu) \le N} \left( \chi_{\mu}^S (\lambda) \right)^2 \right].
\end{align}
This is the character expansion formula for $I_{\mathrm{adj}}^{\mathrm{U}(N)}$.
This formula enables us to calculate the integral \eqref{indexeq} efficiently.
In the following sections, we generalize this formula for $G = \mathrm{USp}(2N)$, $\mathrm{SO}(2N + 1)$, $\mathrm{SO}(2N)$, $\mathrm{O}(2N + 1)$, and $\mathrm{O}(2N)$.

\section{$\mathrm{USp} (2N)$ case}\label{SecUSp}

Let us consider the integral of $G = \mathrm{USp} (2N)$.\ytableausetup{boxsize=0.8em}
We firstly calculate $I_{\mathrm{fund}}^{\mathrm{USp} (2N)}$ using the generalized method of \cite{Dolan:2007rq, Dutta:2007ws}.

We start from \eqref{sumpar}, which holds for an arbitrary $G$. Let $U$ be a $2N \times 2N$ matrix representation of $\mathrm{USp}(2N)$. Its trace is the character of the fundamental representation $\ydiagram[]{1}$:\ytableausetup{boxsize=0.4em} 
\begin{align}
	\mathrm{tr} (U) = \chi_{\ydiagram[]{1}}^{\mathrm{USp} (2N)} (U).
\end{align}
The integral $I_{\mathrm{fund}}^{\mathrm{USp} (2N)}$ is written as
\begin{align}
	I_{\mathrm{fund}}^{\mathrm{USp} (2N)} (x) &= \sum_{\lambda} \left[\frac{ f_{\lambda}(x) }{z_{\lambda}} \int [\diff U]_{\mathrm{USp} (2N)} \prod_{i=1}^{l(\lambda)} \mathrm{tr} \,U^{\lambda_i} \right] .\label{indexspfund}
\end{align}

Now, from the Frobenius formula \eqref{forbeEq} the integrand in equation \eqref{indexspfund} is
\begin{align}
	\prod_{i = 1}^{l(\lambda)} \mathrm{tr} \,U^{\lambda_i} &= t_{\lambda} (U) = \sum_{\mu \vdash |\lambda|: l(\mu) \le 2N} \chi_{\mu}^S (\lambda) \chi_{\mu}^{\mathrm{U}(2N)} (U).
\end{align}
Here, note that the character appearing on the right-hand side is not $\chi_\mu^{\mathrm{USp} (2N)} (U)$ but $\chi_\mu^{\mathrm{U} (2N)} (U)$. As a result, the integral in \eqref{indexspfund} is
\begin{align}
	\int [\diff U]_{\mathrm{USp} (2N)} \prod_{i = 1}^{l(\lambda)} \mathrm{tr} \,U^{\lambda_i} &= \sum_{\mu \vdash |\lambda|: l(\mu) \le 2N} \chi_{\mu}^S (\lambda) \int [\diff U]_{\mathrm{USp} (2N)} \chi_{\mu}^{\mathrm{U}(2N)} (U).\label{frobeCsum}
\end{align}
According to \cite{youngBCD}, $\chi_{\mu}^{\mathrm{U}(2N)} (U)$ can be represented using $\chi_{\lambda}^{\mathrm{USp}(2N)} (U)$ as
\begin{align}
	\chi_{\mu}^{\mathrm{U}(2N)} (U) &= \sum_{\lambda} c_{\mu \lambda}^{\prime} \chi_{\lambda}^{\mathrm{USp}(2N)} (U).\label{decomC}
\end{align}
where $c_{\mu\lambda}^{\prime} \equiv \sum_{\kappa} \mathrm{LR}_{{}^t (2\kappa) \lambda}^{\mu}$.
$\mathrm{LR}_{\nu \lambda}^{\mu}$ are non-negative integers, called the Littlewood-Richardson coefficients.
${}^t (2\kappa)$ is the transposition of $(2 \kappa) = (2\kappa_1, 2 \kappa_2 \dots)$, and ${}^t (2\kappa)$ corresponds to a tensor made from the $\mathrm{USp}(2N)$ invariant antisymmetric tensor $\omega_{ij}$. 

Therefore, using a property of $\mathrm{LR}_{\nu \lambda}^{\mu}$: $\mathrm{LR}_{\nu \bullet}^\mu = \delta_{\nu}^{\mu}$,
where $\bullet$ denotes the trivial representation, 
\begin{align}
	\int [\diff U]_{\mathrm{USp} (2N)} \chi_{\mu}^{\mathrm{U}(2N)} (U) &= \int [\diff U]_{\mathrm{USp} (2N)} \sum_{\lambda} c_{\mu \lambda}^{\prime} \chi_{\lambda}^{\mathrm{USp}(2N)} (U) \notag \\
	&= c_{\mu \bullet}^{\prime}  = \sum_{\kappa} \delta_{{}^t(2\kappa)}^{\mu}
\end{align}
holds.
Thus, the integral on the right-hand side of \eqref{frobeCsum} is one if and only if all $k_m$ of $\mu$ are even, or in Young diagrams, if and only if every column has an even number of boxes, and it is 0 otherwise. In other words,
\begin{align}
	\begin{cases}
		\displaystyle\int [\diff U]_{\mathrm{USp} (2N)} \chi_{\mu}^{\mathrm{U}(2N)} (U) = 1 & \mathrm{for~}\mu \in R_{2N}^{\mathrm{\, c}} (\lvert\mu\rvert) \\
		\displaystyle\int [\diff U]_{\mathrm{USp} (2N)} \chi_{\mu}^{\mathrm{U}(2N)} (U) = 0 & \mathrm{for~}\mu \notin R_{2N}^{\mathrm{\,c}} (\lvert\mu\rvert),
	\end{cases}
\end{align}
where
\begin{align}
	R_{2N}^{\,\mathrm{c}} (|\lambda|) \equiv \left\{ \left.\mu \vdash \lvert\lambda\rvert \; \right|\; l(\mu) \le 2N \; \land \;  \forall m\ (k_m \mathrm{~of~} \mu \mathrm{~is~even})  \right\}.
\end{align}
For example, for $N \ge 3$,\ytableausetup{boxsize=0.8em}
\begin{align}
	R_{2N}^{\,\mathrm{c}} (6) = \left\{\ \ydiagram[]{3,3}\;,\quad\ydiagram[]{2,2,1,1},\;\quad\ydiagram[]{1,1,1,1,1,1} \ \right\}.
\end{align}
Therefore,
\begin{align}
	\int [\diff U]_{\mathrm{USp} (2N)} \prod_{i = 1}^{l(\lambda)} \mathrm{tr} \,U^{\lambda_i} &= \sum_{\mu \in R_{2N}^{\,\mathrm{c}} ( |\lambda|)} \chi_{\mu}^S (\lambda),\label{frobeEqSp}
\end{align}
holds.

By substituting \eqref{frobeEqSp} into \eqref{indexspfund}, we obtain the character expansion formula for $I_{\mathrm{fund}}^{\mathrm{USp} (2N)}$
\begin{align}
	I_{\mathrm{fund}}^{\mathrm{USp} (2N)} (x) &= \sum_{\lambda} \left[ \frac{ f_{\lambda}(x) }{z_{\lambda}}  \sum_{\mu \in R_{2N}^{\,\mathrm{c}} (|\lambda|)} \chi_{\mu}^S (\lambda) \right].\label{indexCf}
\end{align}
In the sum over $\lambda$, we need to include only $\lambda$ with even $|\lambda|$ since $R_{2N}^{\,\mathrm{c}} (|\lambda|) = \varnothing$ when $|\lambda|$ is odd.

In fact, an analytic formula for $I_{\mathrm{fund}}^{\mathrm{USp} (2N)}$ is given in \cite{Balantekin:2001id}.
According to this,
\begin{align}
	\exp\left[ \sum_{m = 1}^{\infty} \frac{1}{m} f(x^m) \mathrm{tr} (U^m) \right] &= \sum_{\lambda\,: l(\lambda) \le N} \underset
{1\le i,j\le N}{\mathrm{det}} \left[ \sum_{p=0}^\infty  A_{p} \left(A_{p+|\lambda_j + i - j|} - A_{p + \lambda_j + 2N + 2 - i - j} \right) \right]\chi_{\lambda}^{\mathrm{USp}(2N)} (U)
\end{align}
holds, where
\begin{align}
	A_{p} = \sum_{\mu \,\vdash\, p} \left[ \frac{f_{\mu} (x)}{z_\mu} \right].
\end{align}
Thus, $I_{\mathrm{fund}}^{\mathrm{USp} (2N)}$ is also represented as
\begin{align}
	I_{\mathrm{fund}}^{\mathrm{USp} (2N)} (x)&= \int [\diff U]_{\mathrm{USp}(2N)} \sum_{\lambda\,: l(\lambda) \le N} \underset
{1\le i,j\le N}{\mathrm{det}}\left[ \sum_{p=0}^\infty  A_{p} \left(A_{p+|\lambda_j + i - j|} - A_{p + \lambda_j + 2N + 2 - i - j} \right) \right]\chi_{\lambda}^{\mathrm{USp}(2N)} (U) \notag\\
		&= \underset
{1\le i,j\le N}{\mathrm{det}}\left[ \sum_{p=0}^\infty  A_{p} \left(A_{p+|i - j|} - A_{p + 2N + 2 - i - j} \right) \right].
\end{align}

Secondly, let us consider $I_{\mathrm{adj}}^{\mathrm{USp} (2N)}$. The adjoint representation of $\mathrm{USp}(2N)$ is the symmetric product \ytableausetup{boxsize=0.8em}$\ydiagram[]{2}$\; of the fundamental representation, and the character is\ytableausetup{boxsize=0.4em}
\begin{align}
	\chi_{\mathrm{adj}}^{\mathrm{USp} (2N)} (U) = \chi_{\ydiagram[]{2}}^{\mathrm{USp} (2N)} (U) &= \frac{1}{2} \left( \left( \mathrm{tr} \,U \right)^2 + \mathrm{tr} \, U^2 \right).
\end{align}
As the result of this equation, the integral $I_{\mathrm{adj}}^{\mathrm{USp} (2N)}$ is written as
\begin{align}
	I_{\mathrm{adj}}^{\mathrm{USp} (2N)} (x) &= \sum_{\lambda} \left[\frac{ f_{\lambda}(x) }{z_{\lambda}} \int [\diff U]_{\mathrm{USp} (2N)} \prod_{i=1}^{l(\lambda)} \left[ \frac{1}{2} \left( \left( \mathrm{tr} \,U^{\lambda_i} \right)^2 + \mathrm{tr} \, U^{2\lambda_i} \right)\right] \right] \notag \\
		&= \sum_{\lambda} \left[\frac{ f_{\lambda}(x) }{z_{\lambda}} \frac{1}{2^{l(\lambda)}} \int [\diff U]_{\mathrm{USp} (2N)}  \sum_{\tilde{\lambda} \in \mathrm{Ev} (\lambda)} \prod_{i = 1}^{l(\tilde{\lambda})} \mathrm{tr} \,U^{\tilde{\lambda}_i} \right].\label{indexsp}
\end{align}
Here, we defined $\mathrm{Ev}(\lambda)$ as follows.
$\mathrm{Ev} (\lambda)$ is the set of partitions obtained by applying one of the following two operations on every part of a given partition $\lambda$.
\begin{itemize}
\item
replace a part $\lambda_i$ of the partition $\lambda$ with $2 \lambda_i$
\item
replace a part $\lambda_i$ of the partition $\lambda$ with two parts $\lambda_i, \lambda_i$
\end{itemize}
Because we have two options for each of $l(\lambda)$ parts,
$\mathrm{Ev}(\lambda)$ has $2^{l(\lambda)}$ parts. \ytableausetup{boxsize=0.8em}
For example, 
\begin{align}
	\mathrm{Ev} \left(\,\ydiagram[]{2,1} \,\right) &= \left\{ \,\ydiagram[]{2,2,1,1}\,,\, \ydiagram[]{4,1,1}\,,\, \ydiagram[]{2,2,2}\,,\, \ydiagram[]{4,2} \, \right\}\\
	\mathrm{Ev} \left(\,\ydiagram[]{2,2}\, \right) &= \left\{\, \ydiagram[]{2,2,2,2}\, ,\, \ydiagram[]{4,2,2}\, ,\, \ydiagram[]{2,2,4} \,\left( =\,\ydiagram[]{4,2,2}\,\right),\, \ydiagram[]{4,4} \, \right\}.
\end{align}
Note that there may be multiple identical diagrams, but they will not be removed.

By substituting \eqref{frobeEqSp} into \eqref{indexsp}, we obtain the character expansion formula of $I_{\mathrm{adj}}^{\mathrm{USp} (2N)}$
\begin{align}
	I_{\mathrm{adj}}^{\mathrm{USp} (2N)} (x) &= \sum_{\lambda} \left[ \frac{ f_{\lambda}(x) }{z_{\lambda}} \frac{1}{2^{l(\lambda)}} \sum_{\tilde{\lambda} \in \mathrm{Ev} (\lambda)} \sum_{\mu \in R_{2N}^{\,\mathrm{c}} (2|\lambda|)} \chi_{\mu}^S (\tilde{\lambda}) \right].\label{indexCa}
\end{align}
Unlike $G = \mathrm{U}(N)$ case, if we want to calculate the contribution of $\lambda$, we need to sum over partitions $\mu$ in $R_{2N}^{\, \mathrm{c}} (2 |\lambda|)$, which has $2 |\lambda|$ boxes.

\section{Orthogonal groups}\label{SecSOO}

Let us consider the special orthogonal groups $G= \mathrm{SO}(n)$ and the orthogonal groups $G=\mathrm{O} (n)$. \ytableausetup{boxsize=0.8em}In both cases, we use the vector representation as ``the fundamental representation'' $\ydiagram[]{1}$.

\subsection{$\mathrm{SO} (n)$ case}\label{SecSO}
Let us firstly consider $I_{\mathrm{fund}}^{\mathrm{SO}(n)}$. 
We again start from \eqref{sumpar}. 
The integral $I_{\mathrm{fund}}^{\mathrm{SO}(n)}$ is written as
\begin{align}
	I_{\mathrm{fund}}^{\mathrm{SO}(n)} (x) &= \sum_{\lambda} \left[\frac{ f_{\lambda}(x) }{z_{\lambda}} \int [\diff U]_{\mathrm{SO} (n)} \prod_{i=1}^{l(\lambda)} \mathrm{tr} \,U^{\lambda_i}\right] \notag \\
		&= \sum_{\lambda} \left[\frac{ f_{\lambda}(x) }{z_{\lambda}}  \int [\diff U]_{\mathrm{SO} (n)}  t_{\lambda} (U) \right].\label{indexSO3}
\end{align}
Next, by the Frobenius formula \eqref{forbeEq},
\begin{align}
	\int [\diff U]_{\mathrm{SO} (n)} t_{\lambda} (U) &= \sum_{\mu \vdash |\lambda|: l(\mu) \le n} \chi_{\mu}^S (\lambda) \int [\diff U]_{\mathrm{SO} (n)} \chi_{\mu}^{\mathrm{U}(n)} (U).\label{frobeEqSO}
\end{align}
According to \cite{youngBCD}, $\chi_{\mu}^{\mathrm{U}(n)} (U)$ can be also represented using $\chi_{\lambda}^{\mathrm{SO}(n)} (U)$ as
\begin{align}
	\chi_{\mu}^{\mathrm{U}(n)} (U) &= \sum_{\lambda} b_{\mu \lambda}^{\prime\prime} \chi_{\lambda}^{\mathrm{SO}(n)} (U).\label{decomB}
\end{align}
where $b_{\mu\lambda}^{\prime\prime} \equiv \displaystyle\sum_{\kappa} \left(\mathrm{LR}_{(2 \kappa) \lambda}^{\mu} + \mathrm{LR}_{(1^n + 2 \kappa) \lambda}^{\mu}\right)$ and $(1^n)$ is the partition $(\underbrace{1,\dots, 1}_{n})$. $(1^n + 2 \kappa)$ means the partition $(1+2\kappa_1,\dots,1+2 \kappa_{l(\kappa)},1, \dots ,1)$.
As well as the $\mathrm{USp}(2N)$ case, $(2\kappa)$ and $(1^n)$ correspond to a tensor made from the $\mathrm{SO}(n)$ invariant symmetric tensor $d_{ij}$ and to the $\mathrm{SO}(n)$ invariant antisymmetric tensor $\varepsilon_{i_1\, i_2 \, \cdots\,  i_n}$: the Levi-Civita tensor\footnote{In the $\mathrm{USp}(2N)$ case, there is also $\varepsilon_{i_1\, i_2 \, \cdots\,  i_{2N}}$ as invariant tensor. However, we do not have to take account of it because $\varepsilon_{i_1\, i_2 \, \cdots\,  i_{2N}}$ can be represented by $\omega_{ij}$.}, respectively.
Therefore,
\begin{align}
	\int [\diff U]_{\mathrm{SO} (n)} \chi_{\mu}^{\mathrm{U}(n)} (U) &= b_{\mu \bullet}^{\prime\prime}  = \sum_{\kappa} \left(\delta_{(2 \kappa)}^{\mu} + \delta_{(1^n + 2 \kappa)}^{\mu} \right)
\end{align}
holds.
Thus, 
the integral on the right-hand side of \eqref{frobeEqSO} is one if and only if all $\mu_i$ are even, or $\mu_{n} \not= 0$ and all $\mu_i$ are odd. 
In other words,
\begin{align}
	\begin{cases}
		\displaystyle\int [\diff U]_{\mathrm{SO} (n)} \chi_{\mu}^{\mathrm{U}(n)} (U) = 1 & \mathrm{for\ }\mu \in R_{n}^{\,\mathrm{r}} (|\mu|) \cup W_{n}^{\,\mathrm{r}} (\lvert\mu \rvert ) \\
		\displaystyle\int [\diff U]_{\mathrm{SO} (n)} \chi_{\mu}^{\mathrm{U}(n)} (U) = 0 & \mathrm{for\ }\mu \notin R_{n}^{\,\mathrm{r}} (|\mu|) \cup W_{n}^{\,\mathrm{r}} (\lvert\mu \rvert )
	\end{cases},\label{SOint}
\end{align}
where
\begin{align}
	R_{n}^\mathrm{\,r} (\lvert \lambda \rvert) &\equiv \left\{ \left.\mu \vdash \lvert\lambda\rvert \;\right|\; l(\mu) \le n \,\land\, \forall i (\mu_i \mathrm{~is~even}) \right\} \\
	W_{n}^\mathrm{\,r} (\lvert \lambda \rvert) &\equiv \left\{ \left.\mu \vdash \lvert\lambda\rvert \;\right|\; l(\mu) = n \,\land\, \forall i (\mu_i \mathrm{~is~odd}) \right\}.
\end{align}
For example in the case of $\mathrm{SO} (4)$,
\begin{align}
	R_{4}^\mathrm{\,r} (6) &= \left\{ \;\ydiagram[]{6}\;,\quad\ydiagram[]{4,2}\;,\quad\ydiagram[]{2,2,2} \; \right\} \\ 
	W_{4}^\mathrm{\,r} (6) &= \left\{ \;\ydiagram[]{3,1,1,1}\; \right\}.
\end{align}
Hence, we obtain the character expansion formula of $I_{\mathrm{fund}}^{\mathrm{SO} (n)}$
\begin{align}
	I_{\mathrm{fund}}^{\mathrm{SO} (n)} (x) &= \sum_{\lambda} \left[ \frac{ f_{\lambda}(x) }{z_{\lambda}}\sum_{\mu \in R_{n}^\mathrm{\,r} (\lvert \lambda \rvert) \cup W_{n}^\mathrm{\,r} (\lvert \lambda \rvert)} \chi_{\mu}^S (\lambda) \right].\label{indexSOf}
\end{align}

As well as $I_{\mathrm{fund}}^{\mathrm{USp}(2N)}$, an analytic formula for $I_{\mathrm{fund}}^{\mathrm{SO} (2N + 1)}$ and $I_{\mathrm{fund}}^{\mathrm{SO} (2N)}$ are given in \cite{Balantekin:2001id}.
Using $A_{p} = \sum_{\mu \,\vdash\, p} \left[ \frac{f_{\mu} (x)}{z_\mu} \right]$, these results are respectively
\begin{align}
	I_{\mathrm{fund}}^{\mathrm{SO} (2N + 1)} (x) &= \sum_{\lambda} \left[ \frac{f_{\lambda} (x)}{z_\lambda} \right]\underset
{1\le i,j\le N}{\mathrm{det}} \left[ \sum_{p=0}^\infty  A_{p} \left(A_{p+|i - j|} - A_{p + 2N + 1 - i - j} \right) \right], \\
	I_{\mathrm{fund}}^{\mathrm{SO} (2N)} (x) &= \frac{1}{2} \underset
{1\le i,j\le N}{\mathrm{det}} \left[ \sum_{p=0}^\infty  A_{p} \left(A_{p+|i - j|} + A_{p + 2N - i - j} \right) \right].
\end{align}

Secondly, let us consider $I_{\mathrm{adj}}^{\mathrm{SO}(n)}$. 
The adjoint representation of $\mathrm{SO}(n)$ is the antisymmetric product \ytableausetup{boxsize=0.8em}$\ydiagram[]{1,1}$\; of the vector representation $\ydiagram[]{1}$\,. The character $\chi_{\mathrm{adj}}^{\mathrm{SO}(n)}$ is given by\ytableausetup{boxsize=0.4em}
\begin{align}
	\chi_{\mathrm{adj}}^{\mathrm{SO} (n)} (U) = \chi_{\ydiagram[]{1,1}}^{\mathrm{SO} (n)} (U) &= \frac{1}{2} \left( \left( \mathrm{tr} \,U \right)^2 - \mathrm{tr} \, U^2 \right).\label{eqtrSO}
\end{align}
From \eqref{sumpar}, the integral $I_{\mathrm{adj}}^{\mathrm{SO}(n)}$ is written as
\begin{align}
	I_{\mathrm{adj}}^{\mathrm{SO}(n)}(x) &= \sum_{\lambda} \left[\frac{ f_{\lambda}(x) }{z_{\lambda}} \int [\diff U]_{\mathrm{SO} (2N + 1)} \prod_{i=1}^{l(\lambda)} \left[ \frac{1}{2} \left( \left( \mathrm{tr} \,U^{\lambda_i} \right)^2 - \mathrm{tr} \, U^{2\lambda_i} \right)\right] \right] \notag \\
		&= \sum_{\lambda} \left[\frac{ f_{\lambda}(x) }{z_{\lambda}} \frac{1}{2^{l(\lambda)}} \int [\diff U]_{\mathrm{SO} (2N + 1)}  \sum_{\tilde{\lambda} \in \mathrm{Ev} (\lambda)} (-1)^{l(\tilde{\lambda})}\prod_{i = 1}^{l(\tilde{\lambda})} \mathrm{tr} \,U^{\tilde{\lambda}_i} \right] \notag \\
		&= \sum_{\lambda} \left[\frac{ f_{\lambda}(x) }{z_{\lambda}} \frac{1}{2^{l(\lambda)}} \int [\diff U]_{\mathrm{SO} (2N + 1)}  \sum_{\tilde{\lambda} \in \mathrm{Ev} (\lambda)} (-1)^{l(\tilde{\lambda})} t_{\tilde\lambda} (U) \right].\label{indexB0}
\end{align}
As well as $I_{\mathrm{fund}}^{\mathrm{SO}(n)}$, using \eqref{frobeEqSO} and \eqref{SOint}, we obtain the character expansion formula of $I_{\mathrm{adj}}^{\mathrm{SO} (n)}$
\begin{align}
	I_{\mathrm{adj}}^{\mathrm{SO} (n)} (x) = \sum_{\lambda} \left[ \frac{ f_{\lambda}(x) }{z_{\lambda}} \frac{1}{2^{l(\lambda)}} \sum_{\tilde{\lambda} \in \mathrm{Ev} (\lambda)} (-1)^{l(\tilde\lambda)}\sum_{\mu \in R_{n}^{\,\mathrm{r}} (2|\lambda|) \cup W_{n}^{\,\mathrm{r}} (2 \lvert\lambda \rvert )} \chi_{\mu}^S (\tilde{\lambda}) \right].\label{indexSOa}
\end{align}

This formula \eqref{indexSOa} holds for all non-negative integers $n$. However, when $n$ is odd integer $2N + 1$, this formula becomes simpler because $W_{2N + 1}^{\,\mathrm{r}} (2 \lvert\lambda \rvert )$ is the empty set. Hence, we also obtain the character expansion formula of $I_{\mathrm{adj}}^{\mathrm{SO} (2N + 1)}$
\begin{align}
	I_{\mathrm{adj}}^{\mathrm{SO} (2N + 1)} (x) = \sum_{\lambda} \left[ \frac{ f_{\lambda}(x) }{z_{\lambda}} \frac{1}{2^{l(\lambda)}} \sum_{\tilde{\lambda} \in \mathrm{Ev} (\lambda)} (-1)^{l(\tilde\lambda)}\sum_{\mu \in R_{2N + 1}^{\,\mathrm{r}} (2|\lambda|)} \chi_{\mu}^S (\tilde{\lambda}) \right].\label{indexBa}
\end{align}
As well as $W_{2N + 1}^{\,\mathrm{r}} (2 \lvert\lambda \rvert ) = \varnothing$, for odd $\lvert\lambda \rvert$, $R_{n}^{\, \mathrm{r}} (|\lambda|) = W_{2N}^{\,\mathrm{r}} (|\lambda|) = \varnothing$ holds.

\subsection{$\mathrm{O}(n)$ case}\label{SecO}
An important difference between $\mathrm{SO}(n)$ and $\mathrm{O}(n)$ is that the Levi-Civita tensor is not an invariant tensor of $\mathrm{O}(n)$ because
\begin{align}
	\varepsilon_{i_1 \cdots i_{2N}} \ \xrightarrow{R \,\in\, \mathrm{O}(2N)}\  &R_{\ \ i_1}^{j_1} \cdots R_{\quad i_{2N}}^{j_{2N}} \varepsilon_{j_1 \cdots j_{2N}} \notag\\
	=\ &\mathrm{det} \left( R \right) \varepsilon_{i_1 \cdots i_{2N}} \notag\\
	=\ &\pm \varepsilon_{i_1 \cdots i_{2N}}.
\end{align}
Therefore, \eqref{SOint} is modified as 
\begin{align}
	\begin{cases}
		\displaystyle\int [\diff U]_{\mathrm{O}(n)} \chi_{\mu}^{\mathrm{U}(n)} = 1 & \mathrm{for\ } \mu \in R_{n}^{\,\mathrm{r}} (|\mu|) \\
		\displaystyle\int [\diff U]_{\mathrm{O}(n)} \chi_{\mu}^{\mathrm{U}(n)} = 0 & \mathrm{for\ } \mu \notin R_{n}^{\,\mathrm{r}} (|\mu|)
		\end{cases}.
\end{align}

Hence, we obtain the character expansion formula of $I_{\mathrm{fund}}^{\mathrm{O} (n)}$
\begin{align}
	I_{\mathrm{fund}}^{\mathrm{O} (n)} (x) &= \sum_{\lambda} \left[ \frac{ f_{\lambda}(x) }{z_{\lambda}}\sum_{\mu \in R_{n}^\mathrm{\,r} (\lvert \lambda \rvert) } \chi_{\mu}^S (\lambda) \right].\label{indexOf}
\end{align}
and that of $I_{\mathrm{adj}}^{\mathrm{O} (n)}$
\begin{align}
	I_{\mathrm{adj}}^{\mathrm{O} (n)} (x) &= \sum_{\lambda} \left[ \frac{ f_{\lambda}(x) }{z_{\lambda}} \frac{1}{2^{l(\lambda)}} \sum_{\tilde{\lambda} \in \mathrm{Ev} (\lambda)} (-1)^{l(\tilde\lambda)}\sum_{\mu \in R_{n}^{\,\mathrm{r}} (2|\lambda|)} \chi_{\mu}^S (\tilde{\lambda}) \right].\label{indexOa}
\end{align}

\section{Large $N$ limit}\label{SecLargeN}

Let us consider the large $N$ (or $n$) limit.
In the character expansion formulas obtained in the previous sections, $N$ (or $n$) appears only through $R_{2N}^{\, \mathrm{c}} (|\tilde\lambda|)$, $R_{n}^{\, \mathrm{r}} (|\tilde\lambda|)$, and $W_{n}^{\, \mathrm{r}} (|\tilde\lambda|)$. The large $N$ limit of $I_R^G (x)$ is obtained by replacing them with their large $N$ limits.
In this paper, we consider only $I_{\mathrm{adj}}^{G}$. 

$I_{\mathrm{adj}}^{\mathrm{SO}(n)}$ and $I_{\mathrm{adj}}^{\mathrm{O}(n)}$ are the same in the large $N$ limit because the large $N$ limit of $R_{n}^{\,\mathrm{r}} (|\tilde{\lambda}|)$ becomes $R_{\infty}^{\,\mathrm{r}} (|\tilde{\lambda}|) \equiv \left\{ \left.\mu \vdash \lvert\tilde\lambda\rvert \;\right|\; \forall i (\mu_i \mathrm{~is~even}) \right\}$ and the large $N$ limit of $W_{n}^\mathrm{\,r} (\lvert \tilde\lambda \rvert)$ contains no partitions with finite size. Thus,
\begin{align}
	I_{\mathrm{adj}}^{\mathrm{SO}(\infty)} (x) = I_{\mathrm{adj}}^{\mathrm{O}(\infty)} (x) &= \sum_{\lambda} \left[ \frac{ f_{\lambda}(x) }{z_{\lambda}} \frac{1}{2^{l(\lambda)}} \sum_{\tilde{\lambda} \in \mathrm{Ev} (\lambda)} (-1)^{l(\tilde\lambda)} \sum_{\mu \in R_{\infty}^{\,\mathrm{r}} (2|\lambda|)} \chi_{\mu}^S (\tilde{\lambda}) \right].\label{indexB2}
\end{align}

Let us consider the $G = \mathrm{USp}(2N)$ case.  The large $N$ limit of $R_{2N}^{\, \mathrm{c}} (|\tilde\lambda|)$ becomes 
$R_{\infty}^{\mathrm{c}} (|\tilde\lambda|) \equiv \left\{ \left.\mu \vdash \lvert\tilde\lambda\rvert \; \right|\; \forall m\, (k_m \mathrm{~of~} \mu \mathrm{~is~even})  \right\}$. Thus, 
\begin{align}
	I_{\mathrm{adj}}^{\mathrm{USp}(\infty)} (x) &= \sum_{\lambda} \left[ \frac{ f_{\lambda}(x) }{z_{\lambda}} \frac{1}{2^{l(\lambda)}} \sum_{\tilde{\lambda} \in \mathrm{Ev} (\lambda)} \sum_{\mu \in R_{\infty}^{\mathrm{c}} (2|\lambda|)} \chi_{\mu}^S (\tilde{\lambda}) \right].\label{indexC2}
\end{align}
We can show that \eqref{indexB2} and \eqref{indexC2} are the same by using the property of the character of symmetric group: $\chi_{\mu}^S (\lambda) = (-1)^{|\lambda| + l(\lambda)} \chi_{{}^t \mu}^S (\lambda)$
and $\mu \in R_{\infty}^{\,\mathrm{c}} (|\tilde\lambda|) \iff {}^t \mu \in R_{\infty}^{\,\mathrm{r}} (|\tilde\lambda|)$. We use $I_{\mathrm{adj}}^{\infty}$ to denote the common large $N$ limit.

Next, for the sum over $R_{\infty}^{\mathrm{c}} (|\tilde\lambda|)$ in \eqref{indexC2}, we claim that the following equation
\begin{align}
	\sum_{\mu \in R_{\infty}^{\mathrm{c}} (|\tilde\lambda|)} \chi_{\mu}^S (\tilde{\lambda}) &= (-1)^{l(\tilde\lambda)}\prod_{m=1}^{\infty} a_{m,k_m}\label{preeq}
\end{align}
holds, where $k_m$ stands for $k_m$ of $\tilde\lambda$: the number of parts with the value $m$ in $\tilde\lambda$ and $a_{m,\,n}$ are defined as
\begin{align}
	\begin{cases}
		a_{m,\,n} = 0 & \mathrm{if\ both\ }m\mathrm{\ and\ } n\mathrm{\ are\ odd} \\
		a_{m,\,n} = m^{n/2} (n-1)!! & \mathrm{if\ }m\mathrm{\ is\ odd\ and\ } n\mathrm{\ is\ even} \\
		a_{m,\,n} = \left.\left(\displaystyle\sqrt{1 + 2m x} \del{}{x} \right)^n e^x \right\rvert_{x=0} & \mathrm{if\ }m\mathrm{\ is\ even}
	\end{cases}.
\end{align}
Furthermore, $a_{m,\,n}$ are the solutions of the following recurrence equations
\begin{align}
	\begin{cases}
		a_{m,\,n} = m (n - 1) a_{m,\,n-2} ,\qquad\quad\quad\; a_{m,\,0} = 1,\; a_{m,\,1} = 0  & \mathrm{if\ }m\mathrm{\ is\ odd} \\
		a_{m,\,n} = a_{m,\,n-1} + m (n - 1) a_{m,\,n-2} ,\quad a_{m,\,0} = a_{m,\,1} = 1  & \mathrm{if\ }m\mathrm{\ is\ even}
\end{cases}.
\end{align}
Although we have not proved \eqref{preeq}, it has been numerically checked for $|\tilde\lambda| \le 20$.
Hence, from \eqref{indexC2} and \eqref{preeq}, the integral $I_{\mathrm{adj}}^{\infty}$ is written as
\begin{align}
	I_{\mathrm{adj}}^{\infty} (x) &= \sum_{\lambda} \left[ \frac{ f_{\lambda}(x) }{z_{\lambda}} \frac{1}{2^{l(\lambda)}} \sum_{\tilde{\lambda} \in \mathrm{Ev} (\lambda)} (-1)^{l(\tilde\lambda)}\prod_{m=1}^{\infty} a_{m,k_m} \right].\label{indexC4}
\end{align}
The infinite product in \eqref{indexC4} is well-defined because $a_{m,k_m} = 1$ for sufficiently large $m$.

\section{Conclusion and Discussion}\label{SecConc}

We gave the formulas for the integral \eqref{indexeq} of $I_{\mathrm{fund}}^{\mathrm{USp}(2N)}$, $I_{\mathrm{adj}}^{\mathrm{USp}(2N)}$, $I_{\mathrm{fund}}^{\mathrm{SO}(n)}$, $I_{\mathrm{adj}}^{\mathrm{SO}(n)}$, $I_{\mathrm{fund}}^{\mathrm{O}(n)}$, and $I_{\mathrm{adj}}^{\mathrm{O}(n)}$.
The results are \eqref{indexCf}, \eqref{indexCa}, \eqref{indexSOf}, \eqref{indexSOa}, \eqref{indexOf}, and \eqref{indexOa}, respectively. They enable us to efficiently calculate the integral \eqref{indexeq} by using the characters of the symmetric group (for partitions with even boxes). Furthermore, we showed that their large $N$ limits are same when $R$ is the adjoint representation, and we proposed the formula \eqref{indexC4}, which give the large $N$ limit without using the characters. 

We can apply the formulas to various systems by setting $f(x)$ appropriately. See Appendix \ref{SecExample} for an  example. 
Furthermore, in this paper, we only consider the integral \eqref{indexeq} with the character of the fundamental representation $\chi_{\mathrm{fund}}^G (U)$ and the character of the adjoint representation $\chi_{\mathrm{adj}}^G (U)$.
It would be nice if we can extend the formulas to the integral with the character of general representations $R$.
For the extension, we need to express $\chi_R^G (U)$ using the representation matrix $U$ of a specific representation.

\section*{Acknowledgements}\label{SecAck}

I would like to express my gratitude to Yosuke Imamura for the suggesting the topic of this research and careful reading of the manuscript.

\newpage
\appendix

\ytableausetup{boxsize=0.8em}

\newpage
\section{Examples}\label{SecExample}
\subsection{The Schur index of 4 dimension $\; \mathscr{N} = 4 \ \mathrm{USp}(2)$ SYM theory}\label{SecExaUSp}
As an example, we will consider the Schur index of $4~\mathrm{dimension} \; \mathrm{USp}(2) \; \mathscr{N} = 4$ SYM theory. That is, $N=1,~x = q, ~f(x) = \frac{2q}{1+q}$, and we will calculate up to the second order of $q$.\ytableausetup{boxsize=0.4em} 

Since the lowest order of $f_{\lambda} (x)$ is $|\lambda|$, the sum of $\lambda$ should be taken for diagrams with box numbers up to 2. Therefore, we can calculate $I_{\mathrm{USp}(2)}(q)$ as
\begin{align}
	I_{\mathrm{USp}(2)}(q) &= 1 + \frac{ f_{\ydiagram[]{1}}(x) }{z_{\ydiagram[]{1}}} \frac{1}{2^{l(\ydiagram[]{1})}} \sum_{\tilde{\lambda} \in \mathrm{Ev} (\ydiagram[]{1})} \sum_{\mu \in R_{2}^{\mathrm{c}} (2|\ydiagram[]{1}|)} \chi_{\mu}^S (\tilde{\lambda}) \notag\newA+ \frac{ f_{\ydiagram[]{2}}(x) }{z_{\ydiagram[]{2}}} \frac{1}{2^{l(\ydiagram[]{2})}} \sum_{\tilde{\lambda} \in \mathrm{Ev} (\ydiagram[]{2})} \sum_{\mu \in R_{2}^{\mathrm{c}} (2|\ydiagram[]{2}|)} \chi_{\mu}^S (\tilde{\lambda}) \notag\newA+ \frac{ f_{\ydiagram[]{1,1}}(x) }{z_{\ydiagram[]{1,1}}} \frac{1}{2^{l\left(\ydiagram[]{1,1} \right)}} \sum_{\tilde{\lambda} \in \mathrm{Ev} \left(\ydiagram[]{1,1} \right)} \sum_{\mu \in R_{2}^{\mathrm{c}} \left(2\left|\ydiagram[]{1,1} \right| \right)} \chi_{\mu}^S (\tilde{\lambda}) \qquad + O(q^3) \notag\\
		&= 1 + \frac{ f(x) }{2} \sum_{\tilde{\lambda} \in \left\{\ydiagram[]{1,1}\;,\;\ydiagram[]{2} \right\}} \sum_{\mu \in \left\{ \ydiagram[]{1,1} \right\}} \chi_{\mu}^S (\tilde{\lambda}) \notag\newA+ \frac{ f(x^2) }{4} \sum_{\tilde{\lambda} \in \left\{\ydiagram[]{2,2}\;,\; \ydiagram[]{4} \right\}} \sum_{\mu \in \left\{ \ydiagram[]{2,2} \right\}} \chi_{\mu}^S (\tilde{\lambda}) \notag\newA+ \frac{ \left(f(x)\right)^2 }{8}  \sum_{\tilde{\lambda} \in \left\{\ydiagram[]{1,1,1,1} \;,\;\ydiagram[]{2,1,1}\;,\;\ydiagram[]{2,1,1}\;,\;\ydiagram[]{2,2} \right\}} \sum_{\mu \in \left\{ \ydiagram[]{2,2} \right\}} \chi_{\mu}^S (\tilde{\lambda}) \qquad + O(q^3) \notag\\
		&= 1 + \frac{ f(x) }{2} (1-1) + \frac{ f(x^2) }{4} (2 + 0) + \frac{ \left(f(x)\right)^2 }{8} (2+0+0+2) + O(q^3) \notag\\
		&= 1 + \frac{ q^2 }{1+q^2} + 2 \frac{ q^2 }{(1+q)^2}  + O(q^3)\notag\\
		&= 1 + 3 q^2 + O(q^3)
\end{align}
using
\begin{align}
	\chi_{\bullet}^S (\bullet) = 1,\, \chi_{\ydiagram[]{1,1}}^S (\ydiagram[]{1,1}) = 1,\, \chi_{\ydiagram[]{1,1}}^S (\ydiagram[]{2}) =- 1,\notag\\ \, \chi_{\ydiagram[]{2,2}}^S (\ydiagram[]{2,2}) = 2,\, \chi_{\ydiagram[]{2,2}}^S (\ydiagram[]{4}) = 0,\, \chi_{\ydiagram[]{2,2}}^S \left(\ydiagram[]{1,1,1,1}\right) = 2,\, \chi_{\ydiagram[]{2,2}}^S \left(\ydiagram[]{2,1,1}\right) = 0.
\end{align}

\end{document}